\documentclass[twocolumn]{aa}
\usepackage{graphicx}

\def\msun{M$_\odot$ }

\begin{document}
   \title{Tracing the X-ray emitting intra-cluster medium of clusters of galaxies beyond $r_{200}$ }

   \author{D. M. Neumann}

   \offprints{D. M. Neumann}

   \institute{AIM - Unit\'e Mixte de Recherche CEA - CNRS - Universit\'e Paris VII - UMR n$^\circ$ 7158 CEA/Saclay; DSM/DAPNIA/Service d'Astrophysique, CEA/Saclay L'Orme des Merisiers Bat. 709, 91191 Gif-sur-Yvette, France}
  
   \date{Received ; accepted }

   \abstract{ We present in this paper a study of a sample of 14 nearby clusters of galaxies ($0.06<z<0.1$) observed with the ROSAT/PSPC. We only select clusters with low galactic absorption (nH$\leq 6\times 10^{20}$cm$^{-2}$) in order to trace the hot X-ray emitting intra-cluster medium (ICM) out to large radii. We convert the X-ray surface brightness profiles of the clusters into emission measure profiles scaled to the classical scaling relations based on the spherical collapse model. We sort the clusters into different temperature bins and stack the scaled emission measure ($ScEM$) profiles of clusters belonging to the same bin together. This method enhances the statistics of the profiles -- especially in outer regions. The stacked profiles allow us to observe a signal out to radii  $r>r_{200}$. 

In the center ($r<0.4 r_{200}$) we find that the $ScEM$ profiles deviate from predicted scaling laws. Hotter clusters have systematically a higher $ScEM$ than cooler clusters. This result is in very good agreement with current studies on the $L_X-T$ relation and the entropy-temperature relation ($S \propto T^{0.65}$) found recently. At radii $r>0.4 r_{200}$ we find that the $ScEM$ profiles agree well within the error bars, suggesting self-similarity. Fitting beta-models to the overall $ScEM$ profiles we find for the different sub samples $r_c=0.15-0.18 r_{200}$ and $\beta=0.8$, which is higher than the canonical value of $\beta=2/3$ often found. The beta-model is generally a better representation for hotter than for cooler clusters. We see indications for continuous steepening of the profiles with increasing radius: at radii $r>0.8 r_{200}$ the profiles are systematically below the beta-model curve with $\beta=0.8$. 

We discuss our results with respect to the observed X-ray luminosity $L_X-T$ relation, the gas mass $M_{gas}-T$ relation and the total mass $M-T$ relation. We also address implications on the origin of the observed $S-T$ relation. Furthermore we discuss the observed steepness of the X-ray profiles, which falls off more rapidly than predicted from the NFW-profile for cold dark matter halos.

   }

   \maketitle
%

\section{Introduction}

In the hierarchical scenario of structure formation, clusters of galaxies are the largest and youngest objects in the universe which have virialized. This makes them ideal targets for cosmological studies. In order to use clusters as probes for cosmology one has to be sure that internal processes did not have time to erase the primordial information from which they have formed and that gravity was the only driving force responsible for the formation of the cluster.

 X-ray observations of clusters of galaxies allow to study cluster physics using the hot intra-cluster medium (ICM), which is spread out all over the cluster (see Sarazin \cite{sarazin} for an introduction). In the past years studies have shown that the central parts of clusters show the presence of non-gravitational physical processes such as cooling flows or cores (e.g. Fabian \cite{f94},\cite{f03} ), interactions with AGN's (e.g. Boehringer et al. \cite{b93} ; Belsole et al. \cite{bsb01}; McNamara et al. \cite{m05}) or sources, which change the ICM temperature from the virial temperature in the central parts (see Voit \cite{v04} for an overview). Therefore, in order to study cosmology with clusters it is wiser to concentrate more in the outer parts of clusters, where internal cluster processes are yet largely inactive due to much lower densities, which require longer timescales. Unfortunately lower density goes together with lower emission: the ICM emits energy mainly via thermal bremsstrahlung, which is proportional to the density squared. Therefore external cluster regions show less X-ray emission with respect to the center, where the density is much higher. This lower emission translates in lower statistics for available X-ray observations. There are several ways to overcome the problem of low statistics in outer regions. One possibility, which is often used is to fit the central parts of clusters with models, typically the beta-model or one of its more sophisticated derivatives, and to extrapolate into outer and undetected regions. The disadvantage of such a method is clear: the extrapolation depends on the validity of the model in external cluster regions. In order to overcome this problem we choose here to go a different way. In order to enhance statistics we employ a stacking method, in which we add different cluster profiles together. Before we add the profiles together we have to make sure that they are similar. For this we apply general cluster scaling laws which have in the past shown to work for observed X-ray cluster profiles (e.g. Mohr et al. \cite{mme99}; Vikhlinin et al. \cite{vfj}; Neumann \& Arnaud \cite{na99}, \cite{na01}; Arnaud et al. \cite{aan}) and translate the X-ray profiles into emission measure profiles, which represent the gas density squares integrated along the line-of-sight. 

Since we want to trace clusters out to large radii we need X-ray observations with large field-of-view and low background. Therefore we choose to look at ROSAT/PSPC (Tr\"umper \cite{rosat}) observations of a sample of nearby clusters.  

The paper is organized as follows. After the introduction we describe briefly the sample selection in Sec.2. In Sec.3 we review the scaling relations which we use for the cluster profiles and the stacking. Sec. 4 describes the data treatment. In Sec.5 we present our results, which we discuss in Sec.6. We finish with the conclusion in Sec.7.

We use the following cosmological parameters through out the paper: $\Omega_{m0}=0.3, \Omega_\Lambda=0.7, H_0=70$~km/s/Mpc. All error bars are 1 $\sigma$.

\section{The sample}

\subsection{General selection criteria}

In order to trace the cluster profiles up to large radii we choose to use ROSAT/PSPC data which offer a large field-of-view of 2 degrees. These data have further the advantage of showing  low background contamination. In order to assure that the cluster emission is well traced out to the virial radius with a large observed region only having background emission, which is important for background subtraction, we restrict ourselves to the redshift interval $0.06<z<0.1$. Furthermore, to avoid high absorption of X-ray emission, which is especially critical in low surface brightness regions, such as the outskirts of clusters, we choose clusters with a hydrogen column density of $nH<6\times 10^{20}$cm$^{-2}$. We only consider clusters with a fitted temperature estimate (preferentially obtained with ASCA or Beppo-SAX). Tab.\ref{tab:sample} shows the clusters in our sample with some of their physical properties. In total we selected 14 clusters for our analysis in the temperature range $1.7< $kT$ <8.5$~keV. 

\subsection{Notes on cluster temperature selection}

When there exist several $kT$ measurements for one cluster we take the latest published measurement. In most of the cases the references come from Ikebe et al. (2002). Comparing the results obtained by these authors with others, specifically with Markevitch et al. \cite{mfs} we find generally good agreement within the error bars.

There exist several $kT$ measurements of A2670, which range from 3.7 (Novicki et al. \cite{nsh}, White, Jones \& Forman \cite{wjf}) up to 5.6~keV (Sanderson et al. \cite{spf}). There are several intermediate measurements which agree within the error bars at around 4.5~keV (Horner et al. \cite{hms}; Finoguenov, David \& Ponman \cite{fdp}), so that we decided finally to include this cluster in our sample at this temperature. There were two clusters, which in principle fulfill our criteria, but which we discarded for further analysis: the first cluster is Abell 2734. Ikebe et al. \cite{irb} fitted kT=$5\pm0.4$~keV for this cluster but Reiprich \& B\"ohringer \cite{rb} estimated kT=$3.9\pm0.6$. These different results are incompatible and we therefore exclude this cluster for our study. The second cluster, which we skipped is RXJ2344. The reason: Ikebe et al. \cite{irb} give large uncertainties of $\Delta$ kT=4~keV, which we consider too high for our analysis.

\begin{table*}
\caption{General cluster properties. For the references: I : Ikebe et al. \cite{irb}; G: Gomez et al. \cite{gomez}; D : De Grandi \& Molendi \cite{dm}; H: Horner et al. \cite{hms}. }             
\label{tab:sample}      
\centering                          
\begin{tabular}{l c c c c c c c }        
\hline\hline                 
name & nH in & z & $kT$ & Ref. & $r_{200}$  & $r_{200}$ & expo. time \\    
  & $10^{20}$ cm$^{-2}$ & & in keV  & & in Mpc & in arcmin & in s \\
\hline                        
A578 & 4.41 & 0.0870 & $1.7^{+1.5}_{-0.3}$ & G & 1.12  & 11.8  &  4519 \\
A644 & 5.14 & 0.0704 & $6.5\pm0.3$ & I & 2.23  & 28.2  &  10246 \\
A1651 & 1.71 & 0.0845 & $6.2\pm0.5$ & I & 2.15 & 23.1  & 7429 \\
A1750 & 2.49 & 0.0852 & $4.5\pm0.3$ & D & 1.83 &  19.6 & 13146 \\
A1775 & 1.00  & 0.0716 & $3.7\pm0.3$ & I & 1.68 & 20.8  & 8669  \\
A1795 & 1.20  & 0.0625 & $6.2\pm0.3$ & I & 2.19 & 30.8  & 62076 \\
A2029 & 3.07 & 0.0773 & $7.9\pm0.4$ & I & 2.45  & 28.4  & 15693 \\
A2142 & 4.05 & 0.0909 & $8.5\pm0.5$ & I & 2.50  & 25.2  & 17215 \\
A2244 & 2.07 & 0.0968 & $5.8\pm0.6$ & I & 2.06  & 19.6  & 2963 \\
A2255 & 2.51 & 0.0806 & $5.9^{+0.4}_{-0.3}$ & I & 2.10 &  23.6 &  14534 \\
A2597 & 2.50 & 0.0852  & $4.2\pm0.5$ & I & 1.77 & 18.9  & 7164 \\
A2670 & 2.92 & 0.0762  & $4.5\pm0.2$ & H & 1.85 & 21.7  & 17679 \\
A3112 & 4.90 & 0.0750 & $4.7\pm0.3$ & I & 1.89 & 22.5 &  7598 \\
A3921 & 2.80 & 0.0936  & $4.9\pm0.4$ & I &1.90 & 18.7  &  11997 \\
\hline                                   
\end{tabular}
\end{table*}

\section{Scaling relations}

\subsection{Virial radius $R$ and $r_{200}$}
\label{sec:r200}

Assuming spherical symmetry the mass of the virialized part of a cluster is related to its virial radius by 
\begin{equation}
M/R^3=\rho_{vir}
\end{equation}

(e.g. Kaiser \cite{kaiser})
Defining $\Delta_c$ as the density contrast with $\Delta_c = \rho_{vir}/\rho_{crit}$; -- $\rho_{crit}=3H_0^2/(8 \pi G)$ is the critical density of the universe and $E^2(z)=\rho_{crit}(z)/\rho_{crit}(z=0)$ we can write:

\begin{equation}
M/R^3=\Delta_c \rho_{crit}(z) \propto \Delta_c  \frac{\rho_{crit}(z)}{\rho_{crit}(z=0)}= \Delta_c E^2(z)
\label{equ:delta}
\end{equation}

Applying the virial theorem to the ICM gives $M \propto RT$, where $T$ is the ICM temperature. This yields together with eq.\ref{equ:delta} the mass temperature relation 
\begin{equation}
M \propto \frac{T^{3/2}}{\sqrt{\Delta_c}E(z)}
\end{equation}

and the radius temperature relation 

\begin{equation}
R \propto \frac{T^{1/2}}{\sqrt{\Delta_c}E(z)}
\label{equ:relrt}
\end{equation}

with 

\begin{equation}
E^2(z)=\Omega_0(1+z)^3+\Omega_\Lambda
\end{equation}

in our choice of cosmological parameters and

\begin{equation}
\Delta_c= 18 \pi^2 +82 (\Omega(z)-1) - 39 (\Omega(z)-1) ^2
\end{equation}

with 

\begin{equation}
\Omega(z)= \frac{\Omega_0 (1+z)^3}{E^2(z)}
\end{equation}

using the usual spherical top hat model (Eke et al. \cite{eke96}; Bryan \& Norman \cite{bn98}). In the case of a critical density universe, $\Delta_c$ is constant with $18 \pi^2$.
 
For the normalization of the x-axis of the surface brightness profiles we use $r_{200}$, the radius in which the mean over-density of the cluster is 200 times the critical density of the universe. Numerical simulations have shown that within this density contrast clusters can be assumed to be in equilibrium (see Cole \& Lacey \cite{cl96}). At larger radii in-fall becomes important and the assumption of virialization is not anymore valid. To calculate $r_{200}$ we use its definition:

\begin{equation}
M_{200} = \frac{4 \pi}{3} 200 \rho_{crit} r_{200}^3
\label{equ:m200}
\end{equation}

We are interested in the $r_{200}$-T relationship at the mean redshift of our sample, z=0.08. With the choice of our cosmological parameters we find $\rho_{crit}$ = 146 \msun /kpc$^3$. To calculate $r_{200}$ we need a relationship between $M$ and $r$.To establish this relationship we can use the hydrostatic equation using the X-ray emitting intra-cluster medium (ICM):

\begin{equation}
\frac{1}{\rho_g}\frac{\mbox{d}P}{\mbox{d}r} = -\frac{\mbox{d}\Phi}{\mbox{d}r} = -\frac{GM(r)}{r^2}
\label{equ:hydro}
\end{equation}

$\rho_g$ is the gas density and $P$ is the pressure. We can use the equation of state of a perfect gas for the ICM: $P=nkT$. Assuming that the ICM is isothermal we only need the gas density distribution of the gas to calculate the total mass at a given radius $r$. Referring to the beta-model, which we will introduce in Sec.\ref{sec:beta} and more specifically in eq.\ref{equ:beta} we can rewrite eq.\ref{equ:hydro} as:

\begin{equation}
M(r) = \frac{3 k \beta}{G \mu m_p} \frac{T r^3}{r^2 + r_c^2}
\end{equation}

$\mu=0.59$ is the mean molecular weight. We find for $\beta$ values of 0.80 (see Sec.\ref{sec:beta}). Since we are interested in the outer parts of clusters, where $r>>r_c$, we can neglect $r_c$ for getting $r_{200}$.

Therefore we can approximate at large radii:

\begin{equation}
M(r) = \frac{3 k 0.8 T r}{G \mu m_p}
\label{equ:mlarge}
\end{equation}

Inserting eq.\ref{equ:mlarge} into eq.\ref{equ:m200} gives:

\begin{equation}
r_{200} = 0.86 \sqrt{\frac{kT}{\mbox{keV}}} \mbox{Mpc}
\label{equ:r200}
\end{equation}

at $z=0.08$, which is our $r_{200}$-T relation at this redshift. We will discuss this and the resulting $M-T$ relation in detail in the discussion. 

Now, practically, to scale the profiles to $r_{200}$ we do the following: we calculate for each cluster the angular diameter distance and translate the surface brightness profile into a profile with Mpc scale using our adopted cosmological parameters. We then apply the relationship shown in eq.\ref{equ:relrt} for each cluster to place each cluster at z=0.08.- We multiply the radius in Mpc scale by $(\sqrt{\Delta_c(z=0.08)}E(z=0.08))/(\sqrt{\Delta_c(z)}E(z))$. Please note: the redshift interval is very small so that this correction is not very important. Now, that we ``virtually red-shifted'' the cluster to z=0.08 we can apply eq.\ref{equ:r200} to calculate its $r_{200}$ and divide the Mpc scale of the surface brightness profile by this value. The surface brightness profile is now in units of $r_{200}$. In Tab.\ref{tab:sample} we display the physical parameters which are necessary to calculate $r_{200}$ for each cluster.

\subsection{Emission measure profiles and self-similarity}

The emission measure is defined to be the gas density squared integrated along the line-of-sight:

\begin{equation}
EM(r) = \int  n_g^2(l) dl = \int_r^R \frac{n_g^2(x)xdx}{\sqrt{x^2-r^2}}
\end{equation}

Assuming self-similarity, the mass-temperature and radius-temperature relation and supposing that $M_{gas} \propto M$ we find: 

\begin{equation}
EM(r) \propto \frac{M^2}{R^5} \propto \sqrt{T}  (\sqrt{\Delta_c}E(z))^3
\label{equ:emrsc}
\end{equation}

however, $EM(r)$ is not measured directly. What can be observed is the X-ray surface brightness of ICM (see below), which is 

\begin{equation}
S(r)=\Lambda(T,z) \frac{EM(r)}{4\pi (1+z)^4}
\end{equation}

$\Lambda(T,z)$ is in our case the emissivity in the chosen ROSAT band (0.5-2.0~keV) taking into account interstellar absorption and the PSPC spectral response. $\Lambda(T,z)$ can be calculated by simulating a red-shifted cluster spectrum with a given plasma temperature and folding it through the spectral response of the ROSAT/PSPC. In our case we use XSPEC for the spectral simulations. One can thus translate the observed surface brightness profile into the emission measure $EM(r)$ profile. Self-similarity of clusters implies that the scaled emission measure profiles should be identical for all clusters:  

\begin{equation}
ScEM(r/r_{200}) = \frac{4 \pi (1+z)^4 S(r/r_{200})}{\Lambda(T,z)\sqrt{T} (\sqrt{\Delta_c} E(z))^3} 
\label{equ:scem}
\end{equation}

$ScEM(r/r_{200})$ can be calculated directly from the observed surface brightness distribution. We apply eq.\ref{equ:scem} for all the cluster profiles using the best fit temperature displayed in Tab.\ref{tab:sample}

\section{Data treatment}

\subsection{Surface brightness profile}

For each cluster we create an image in the 0.5-2.0~keV range and calculate the corresponding exposure maps taking into account wobbling of the satellite and the geometry of the support structure of the PSPC. We create from these images  radial surface brightness profiles with a bin size of each annulus of 15~arcsec. Obvious point sources or clearly identified substructures are cut out. All profiles have 220 bins (or 55~arcmin) with the exception of A1775: this cluster was observed off-axis and we have to restrict ourselves to 190 bins in this case. 

We use the software package EXSAS for these tasks

\subsection{Background subtraction}

Since we are interested in the profiles of the outskirts of the clusters where source emission is low, background subtraction is a crucial point in our analysis.

To determine the background for each cluster we calculate the mean emission $B_1$  in the radial range $1.5<r/r_{200}<2.5$. In this region we are confident to not anymore detect cluster emission. In order to check whether the background is flat we also calculate the background in the region  $1.5<r/r_{200}<2$ ($B_2$). If $|B_1 - 2\times \sigma/\sqrt{no.}|< B_2$, where $\sigma$ is the standard deviation of $B_2$ and $no.$ is the number of radial bins which fulfill the criterion  $1.5<r/r_{200}<2$, we consider the background emission as flat. There are some clusters (A644, A2029, and A1795) for which our profile does not go up to $2\times r_{200}$. In this case we consider for $B_1$ the region in which $r/r_{200}>1.$ and for $B_2$ $1.0<r/r_{200}<1.5$. All clusters fulfill the criterion of $|B_1 - 2\times \sigma/\sqrt{no.}|< B_2$ with the exception of A3112 and A2670. In these particular cases we determine the background $B_1$ in the interval $1.0<r/r_{200}<1.5$. For all clusters we use $B_1$ as background estimate in the following, which we subtract from the original extracted surface brightness profile. The error of the background estimates is typically 10\% of the statistical error in the surface brightness bins close to $r_{200}$. Adding this background uncertainty quadratically to the statisical error would enhance the error by roughly 1\%, which is negligible. In the following we therefore do not take into account the error of the background. 

We applied several methods to determine the background level and checked the robustness of our results with different background subtractions. All background subtractions we performed yielded the same results.

\subsection{From the surface brightness profiles to the scaled and stacked emission measure (ScEM) profiles}

To obtain the $ScEM$ profile for each cluster we use the background subtracted surface brightness profiles and apply eq.\ref{equ:scem} to the y-axis. The scaling to $r_{200}$ is explained in Sec.\ref{sec:r200}.

\subsection{Stacking and sub sample selection}

\begin{figure}
\resizebox{\hsize}{!}{\includegraphics{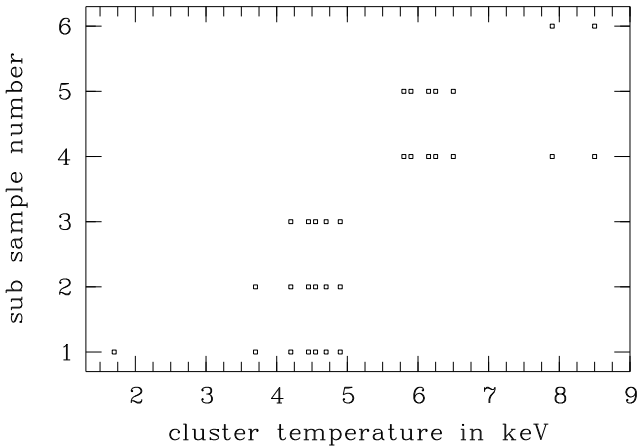}}
\caption{The temperature distribution of the clusters (see also Tab.\ref{tab:sample}) in our sample and the selection in temperature of sub sample 1 to 6.}
\label{fig:kt}
\end{figure}

We are interested in the outer parts of clusters, where statistics is generally sparse. In order to improve the statistics we opt for a stacking method, in which we add different emission measure profiles together. We also want to study the temperature dependence of the scaled emission measure profile. Therefore we divide our sample up in sub samples which represent different cluster temperatures. Fig.\ref{fig:kt} shows the temperature distribution in our sample and the selection of the 6 sub samples in different temperature intervals. The sub sample composition is also shown in Tab.\ref{tab:kt}.

\begin{figure}
\resizebox{\hsize}{!}{\includegraphics{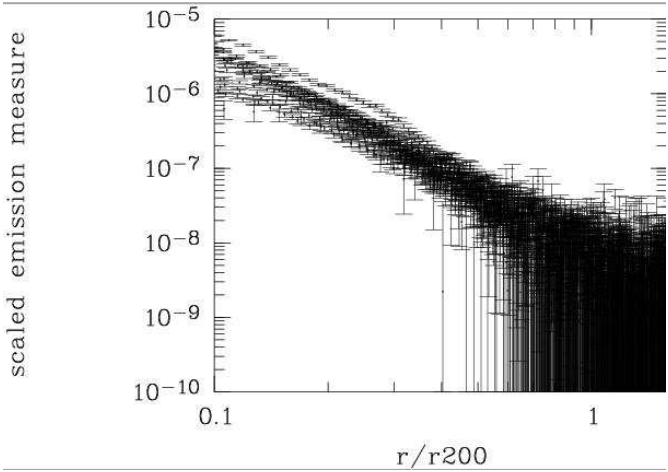}}
\caption{The scaled emission measure profiles of all clusters before stacking. The error bars here are based on pure photon statistics.}
\label{fig:scemall}
\end{figure}

\begin{table*}
\caption{The membership list of clusters of the different sub samples. The first column shows the sub sample number, the second column the temperature interval for each sub sample and the third column shows the average temperature of each sub sample.}
\label{tab:kt}
\begin{tabular}{l c c c c c c c c c}
\hline\hline 
no. &  $kT$   &   $\overline{kT}$  & name & ... &  ... & ... &  ...& ... & ...  \\
& min-max & [keV] & ([keV]) & & & & & & \\
 & [keV] &   &  &  &  & & & & \\
\hline
1 & 1.7-4.9 & 4.0 & A578 (1.7) & A1775 (3.7) & A2597 (4.2) & A1750 (4.5) & A2670 (4.5) & A3112 (4.7) & A3921 (4.9)\\
2 & 3.7-4.9  & 4.4  & &  A1775 (3.7) & A2597 (4.2) & A1750 (4.5) & A2670 (4.5) & A3112 (4.7) & A3921 (4.9)  \\
3 & 4.2-4.9 & 4.6  & & & A2597 (4.2) & A1750 (4.5) & A2670 (4.5) & A3112 (4.7) & A3921 (4.9) \\
\hline \hline
4 & 5.8-8.5 & 6.7  & A2244 (5.8) & A2255 (5.9) & A1795 (6.2) & A1651 (6.2) & A644 (6.5) & A2029 (7.9) & A2142 (8.5) \\
5 & 5.8-6.5 & 6.1  & A2244 (5.8) & A2255 (5.9) & A1795 (6.2) & A1651 (6.2) & A644 (6.5) & & \\
6 & 7.9-8.5 & 8.2 &  & & & & & A2029 (7.9) & A2142 (8.5) \\
\hline
\end{tabular}
\end{table*}

Fig.\ref{fig:scemall} shows the scaled emission measure profiles for all clusters before stacking. 

In order to stack the scaled profiles together all the points of the different $ScEM$ profiles must be located at the same relative position with respect to $r_{200}$, the radius to which we scale the profiles. For this we interpolate all scaled emission measure profiles linearly with a step-width of $0.03\times r_{200}$. This gives for each profile 33 points up to the virial radius. This is about half of the points we have for each original emission measure profile. With the reduction of the data points we make sure that each original data point is only used once for the linear interpolation and that the resulting data points are thus uncorrelated. We calculate the mean scaled emission measure for each sub sample:

\begin{equation}
\overline{ScEM_k}(r/r_{200}) = \frac{\sum_{i=j}^{n} ScEM(i,r_{200})}{n-j+1}
\end{equation}

$k$ is the index for the sub sample and ranges from 1 to 6 (see also Tab.\ref{tab:kt}). $j$ and $n$ is the number of the first and of the last cluster in the sub sample. The number of the cluster raises with temperature, so the coldest cluster, A578 has number 1 and the hottest cluster, A2142 has number 14 (see also Tab.\ref{tab:kt}). We also calculate $\overline{ScEM}(r)$ for all 14 clusters together.

\subsubsection{The uncertainty of the scaled emission measure profiles}
\label{sec:uncer}

There are different possibilities to calculate the uncertainty of the scaled and stacked emission measure profiles. As a first attempt one can add quadratically the different statistical errors of the individual profiles based on photon numbers. 

\begin{equation}
\Delta_{stat} \overline{ScEM_k}(r/r_{200}) = \frac{\sqrt{ \sum_{i=j}^{n} (\Delta_{stat} ScEM(i,\frac{r}{r_{200}}))^2}}{n-j+1} 
\end{equation}

$\Delta_{stat} \overline{ScEM}(r/r_{200})$ is the mean error of the stacked profile taking into account the internal error of each profile. However, in this error calculation differences linked to effects such as morphology (substructure, ellipticity), redshift or temperature uncertainties are not taken into account. Another approach is to calculate the dispersion for the stacked profiles at a given radius and to determine the mean error from this. The mean error per point for the stacked profiles is in this case:

\begin{equation}
\Delta \overline{ScEM}(r/r_{200}) = \frac{\sigma(r/r_{200})}{\sqrt{n-j+1}}
\label{equ:err}
\end{equation}

where $\sigma(r/r_{200})$ is the rms at $r/r_{200}$ averaged over all profiles in the sub sample. 

If only uncertainties due to photon statistics were important $\Delta_{stat} \overline{ScEM}(i)$ and $\Delta \overline{ScEM}(i)$ should be of the same order. Else, $\Delta \overline{ScEM}(i)$ has higher values than $\Delta_{stat} \overline{ScEM}(i)$. Therefore, to be conservative we choose in the following $\Delta \overline{ScEM}(i)$ for the error bars. As we will see, in the central parts there is a systematic trend of higher $ScEM$ with higher temperature. This increases the dispersion of the scaled emission measure profiles for each sub sample and is also, of course, dependent on the temperature interval of the sub sample. The smaller the temperature interval, the smaller should be the apparent dispersion. This effect is not of statistical origin but is systematics. This results in a likely overestimate of the calculated error bars in the central parts.

\section{Results}

\subsection{The different stacked emission measure profiles}

\begin{figure}
\resizebox{\hsize}{!}{\includegraphics{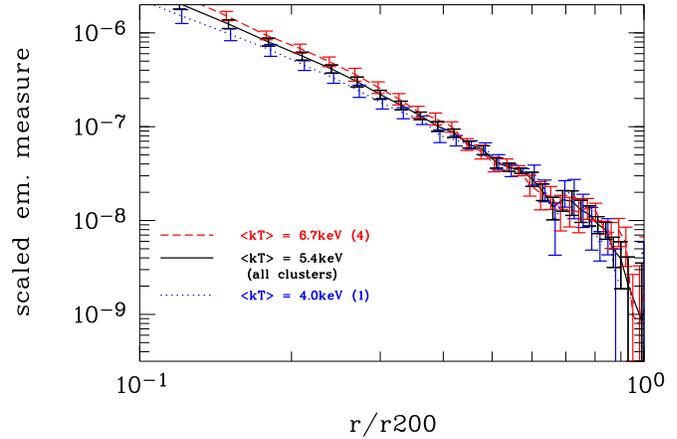}}
\caption{The stacked scaled emission measure profile of sub sample 1 and 4 and for all clusters together.}
\label{fig:disp14}
\end{figure}

\begin{figure}
\resizebox{\hsize}{!}{\includegraphics{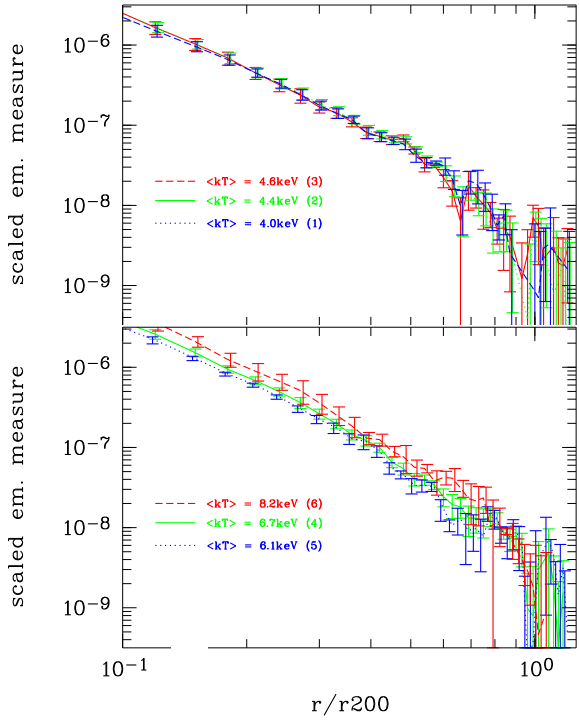}}
\caption{The mean scaled emission measure profile and the mean profiles for sub sample 1, 2 and 3 (top panel) and sub sample 4, 5, and 6 (bottom panel).Please note: sub sample 6 comprises only 2 clusters and the shown error based on the dispersion of the profiles has to be interpreted very cautiously.}
\label{fig:dispsub}
\end{figure}

Fig.\ref{fig:disp14} shows the scaled emission measure profile averaged over  all clusters as well as the mean profile of sub samples 1 and 4 (see also Tab.\ref{tab:kt}), which each represent half of all clusters. Up to about $0.4 r_{200}$ we see a systematic trend of higher $ScEM$ for higher $kT$ clusters, which signifies a deviation from self-similarity and implies that apart from the spherical collapse based on gravity, additional clusters physics plays a role in these regions. At larger radii the different $ScEM$ profiles agree well within the error bars. In Fig.\ref{fig:dispsub} we show the stacked emission measure profiles of the smaller sub samples. Also here, in the central parts the same trend is visible. The higher the average cluster temperature the higher the central values for the $ScEM$ profile. The profiles start to agree within the error bars at different radii. This effect might be at least partially caused by the low number of clusters in the sub samples. Sub Sample 6 does not allow a proper assessment of the statistical uncertainties since it only contains 2 clusters. Additionally, one of the clusters in this sub sample is Abell 2142, which is known to contain an important amount of substructure (Markevitch et al. \cite{a2142}).

\begin{figure}
\resizebox{\hsize}{!}{\includegraphics{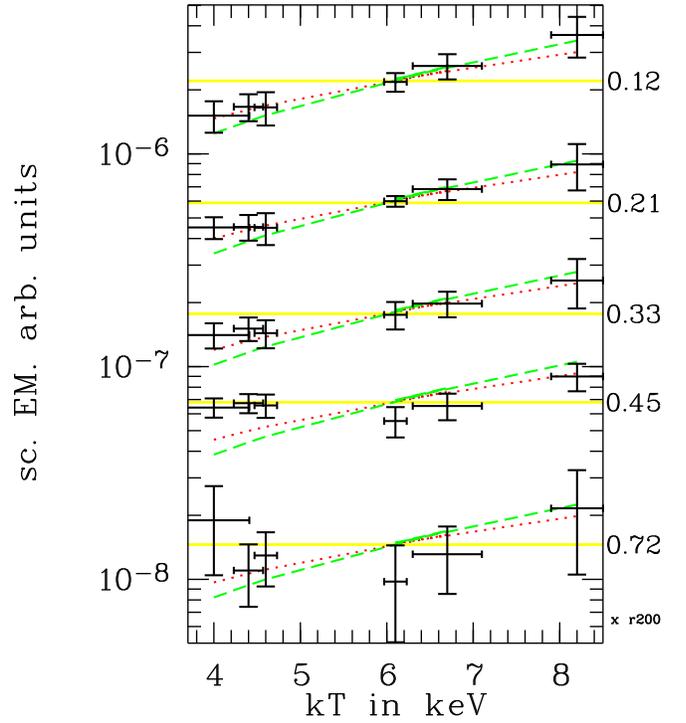}}
\caption{The scaled emission measure as a function of temperature for different radii. The corresponding radii are shown in units of $r_{200}$ right to the figure. The error bars in the x-direction show the statistical uncertainty of the cluster temperature in each sub sample (see text). The horizontal lines show the mean $ScEM$ averaged over all the sub samples at the selected radii. The dotted line shows the relationship $ScEM\propto T$ and the dashed line $ScEM \propto T^{1.4}$. Please note: sub sample 6 comprises only 2 clusters and the shown error based on the dispersion of the profiles has to be interpreted cautiously.}
\label{fig:disptemp}
\end{figure}

We show in Fig.\ref{fig:disptemp} the scaled emission measure at different radii as a function of average sub sample temperature (see Tab.\ref{tab:kt}). Displayed is at the same time the rms of the best fit cluster temperatures divided by the square root of the number of clusters in each sub sample (error bars in x-direction). In Fig.\ref{fig:disptemp} we see again up to about $~0.4r_{200}$ a systematic trend of higher $ScEM$ for hotter clusters, which disappears, when approaching $~0.4 r_{200}$. We will discuss the differences of the $ScEM$ profiles in the central parts in Sec.\ref{sec:disc}.

\subsection{Quantification of differences}

In order to quantify the differences between the different profiles we apply a $\chi^2$-test in different radial intervals with $r_1 \leq r \leq r_2$ , in which $\chi^2$ is defined in the following way:

\begin{equation}
\chi^2 = \sum_{r_1}^{r_2}\frac{(\overline{ScEM_1}(r_i)-\overline{ScEM_4(r_i)})^2}{\Delta \overline{ScEM_1(r_i)}^2 + \Delta \overline{ScEM_4(r_i)}^2}
\end{equation}

$\Delta \overline{ScEM_i}$ is defined in eq.\ref{equ:err}. We only concentrate here on sub sample 1 and 4 since they are the only ones in which we have enough clusters to achieve sufficiently good statistics for this test. The corresponding results are shown in Tab.\ref{tab:chi2}. The $\chi^2$ results vary from 8.5 for the central parts down to 1.2 in the external parts of the clusters. The lower the value of $\chi^2$ the higher is the probability that the 2 profiles have the same shape and normalization. The lower values for the external regions are explained by the larger error bars and the apparent missing trend that hotter profiles show a higher scaled emission measure. It is a reconfirmation that the profiles do not coincide in the central parts.

\begin{table}
\caption{$\chi^2$ test results between sub sample 1 and 4 in different radial regimes. The first two columns give the inner and the outer boundary of the radial interval taken for the test.}
\label{tab:chi2}
\begin{tabular}{ccccc}
\hline\hline
int. radius & ext. radius & $\chi^2$ & d.o.f. & $\chi^2$ / d.o.f. \\
in $r_{200}$ & in $r_{200}$ \\
\hline
0.1 & 0.2 & 16.9 & 2 & 8.45 \\
0.1 & 0.3 & 35.4 & 6 & 5.9 \\
0.1 & 0.4 & 43.3 & 9 & 4.81 \\
0.1 & 0.7 & 50.0 & 19 & 2.61\\
0.1 & 1.0 & 67,5 & 29 & 2.33 \\
0.2 & 0.3 & 18.5 & 3 & 6.17 \\
0.2 & 0.4 & 26.4 & 6 & 4.4 \\
0.3 & 1.0 & 32.0 & 22 & 1.46 \\
0.4 & 1.0 & 24.1 & 19 & 1.27 \\
0.6 & 1.2 & 22.8 & 19 & 1.20 \\
0.7 & 1.2 & 20.8 & 16 & 1.30 \\
0.8 & 1.2 & 19.8 & 13 & 1.52 \\ 
\hline
\end{tabular}
\end{table}

\subsection{Beta-model fits}

\label{sec:beta}

The beta-model is defined the following way see (Cavaliere \& Fusco-Femiano \cite{cf76}):

\begin{equation}
S(r) = S_0 (1+\frac{r^2}{r_c^2})^{-3\beta+0.5} 
\end{equation}

$S(r)$ is the surface brightness. $S_0$ is the central surface brightness, $r_c$ the core radius and $\beta$ a slope parameter. The beta-model is such a successful model since it allows to deproject easily the surface brightness of the ICM which emits via thermal bremsstrahlung into a gas density profile:

\begin{equation}
n(r) = n_0 (1+\frac{r^2}{r_c^2})^{-3\beta/2} 
\label{equ:beta}
\end{equation}

Emission measure profiles have the same shape as surface brightness profiles. Therefore we can use the beta-model also for our stacked emission measure profiles. To further quantify the differences of the various scaled emission measure profiles we fit beta-models in different radial intervals. Again, for statistical reasons we only concentrate here on sub sample 1 and 4. Tab.\ref{tab:beta} resumes our results. We fit in 4 different radial intervals. We do not look at regions $<0.1\times r_{200}$ since central parts of clusters are often affected by so-called cold cores, which raise the central surface brightness of some clusters and which subsequently enhance the dispersion of the profiles in these regions. 

\begin{figure}
\resizebox{\hsize}{!}{\includegraphics{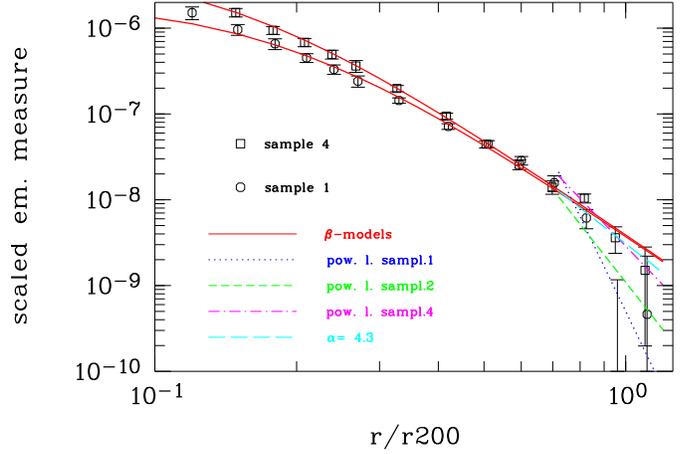}}
\caption{The profiles of sub sample 1 and 4 with logarithmic binning  with different models superposed. The curve corresponding to $\alpha=4.3$ is our lower limit for the slope in the cluster outskirts (see text). This slope matches at the same time a density profile which follows the NFW-profile at these radii (see text). We normalize this $\alpha=4.3$ profile to match the data points at $r=0.7 r_{200}$.}
\label{fig:beta}
\end{figure}

\begin{figure}
\resizebox{\hsize}{!}{\includegraphics{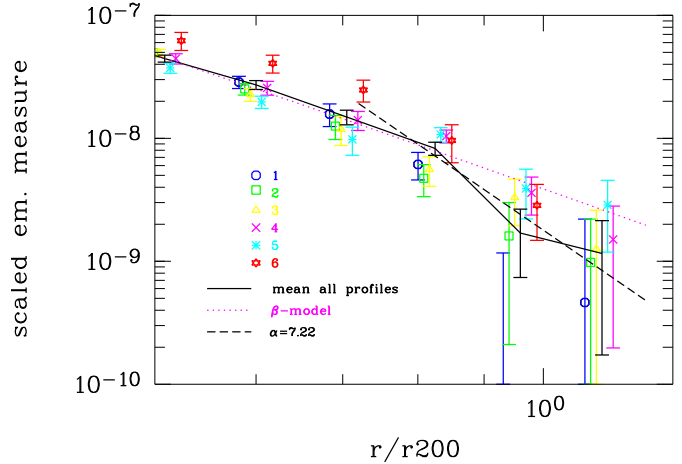}}
\caption{Zoom into the outskirts of the ScEM-profiles. The curve showing a power law with slope $\alpha=7.22$ is based on the fit in the outer parts of all $ScEM$ profiles stacked together (see also Tab.\ref{tab:power}). The symbols and error bars of the different sub samples are shifted with respect to each other only for visibility. The points of the different profiles have all the same location in x-direction.}
\label{fig:out}
\end{figure}

As can be seen in Tab.\ref{tab:beta} for the fit in the different regions the parameters $\beta$ and $r_c$ change less for sub sample 4 (the hotter clusters) than for sub sample 1 (the cooler clusters). This indicates that the beta-model is a better description for hotter than for cooler clusters. In Fig.\ref{fig:beta} we show the $ScEM$ profiles with logarithmic binning and the best fit beta-model profiles for $0.1<r/r_{200}<1.2$ . We see that the beta-model is too shallow for sample 1 in the center and at radii $r>0.8r_{200}$. For sample 4 we see that the beta-model is a good description in the center, however, also in this sample we see indications that the profile falls off more rapidly in the outskirts than predicted with our  best fit overall beta-model. In Fig.\ref{fig:out} we zoom into the external parts of the $ScEM$ profile and show the results of the different sub samples with logarithmic binning. There is an obvious trend that the data points lie below the beta-model curves at radii $r>0.8 r_{200}$. 

In Fig.\ref{fig:beta} we see that the fitted beta-models for sub sample 1 and 4 are identical at radii $r>0.6 r_{200}$. This can be easily explained by the fact that we fit identical $\beta$'s for the 2 samples. The differences found for the core radii do not play anymore an important role in the outskirts of clusters.

The found $\chi^2_{red}$ values are systematically lower for sub sample 4 than for sub sample 1. This again indicates that the beta-model is a better description for hotter than for cooler clusters.

For the two samples we find a mean $\beta$ of about $\overline{\beta} = 0.8$ and  $\overline{r_c} = 0.15 r_{200}$ for sub sample 4 and   $\overline{r_c} = 0.19r_{200}$ for sub sample 1.  The mean value for $\beta$ which we find is with around 0.8 higher than the canonical value of 2/3 as often found by fitting individual clusters without any stacking. 

If we perform beta-model fits with $\beta=2/3$ fixed, we obtain values of $r_c$ of $0.09 r_{200}$ for sub sample 1 (the cooler clusters) and $0.039 r_{200}$ for the hotter ones. The corresponding reduced $\chi^2$ values are 1.9 (sub sample 1) and 1.6 (sub sample 4). These values are significantly higher than the reduced $\chi^2$ values for our best fits leaving $\beta$ as a free parameter. 

\begin{table}
\caption{Results of the beta-model fitting. Error bars are $1\sigma$-level. The first column shows the sub sample number. The second shows the interval in units of $r_{200}$ in which the beta-model is fitted. $r_c$ is also in units of $r_{200}$.}
\label{tab:beta}
\begin{tabular}{cccccccc}
\hline\hline
sa. & $r_{min}-$ & $r_c$ & $\beta$ & $S_0\times$ & d.o.f. & $\chi^2_{red.}$ \\
& $r_{max}$ & & &  $ 10^{-6}$ & & \\
\hline
1 & 0.1-1.2  & $0.19^{0.21}_{0.18}$ & $0.80^{0.84}_{0.77}$ & 2.1 & 33 & 1.42 \\
1 & 0.1-1.0 & $0.18^{0.19}_{0.17}$ & $0.78^{0.82}_{0.75} $ & 2.3 & 26 & 1.66 \\
1 & 0.1-0.7 & $0.089^{0.094}_{0.082} $ & $0.64^{0.66}_{0.62}$ & 6.6 & 16 & 0.48\\
1 & 0.3-1.2 & $1.4^{1.7}_{0.94}$ & $4.30^{6.46}_{2.37} $ & 2.5 & 26 & 0.72 \\
\hline
4 & 0.1-1.2 & $0.15^{0.16}_{0.14}$ & $0.80^{0.82}_{0.78}$ & 5.8 & 33 & 0.68 \\
4 & 0.1-1.0 & $0.136^{0.143}_{0.130}$ & $0.78^{0.80}_{0.75}$ & 6.5 & 26 & 0.61 \\
4 & 0.1-0.7 & $0.13^{0.14}_{0.12}$ & $0.77^{0.80}_{0.74}$ & 6.9 & 16 & 0.30 \\
4 & 0.3-1.2 & $0.25^{0.27}_{0.22}$ & $0.89^{0.95}_{0.84}$ & 1.8 & 26 & 0.88 \\ 
\hline
\end{tabular}
\end{table}

\subsection{Power law fits}

Beside the beta-model fits we also perform power law fits in different radial intervals for sub sample 1 and 4. The power law we fit has the following form:

\begin{equation}
S(r)= N_{200} r^{-\alpha}
\label{equ:power}
\end{equation}

$N_{200}$ is the normalization of the power law at $r_{200}$ and $\alpha$ is the power law index. Our fit results are displayed in Tab.\ref{tab:power}. We fit power laws with increasing $\alpha$ as we go to larger and larger radii, which shows a continuous steepening of the profiles. In the outskirts of the cluster a beta-model with $\beta=0.8$, as found in the previous section gives $\alpha=3.8$. The power law fits we find for radii $>0.7 r_{200}$ are systematically higher than this. This indicates again that the $ScEM$ profiles show a steeper slope in the external parts of the cluster than predicted by a beta-model with $\beta=0.8$. $\alpha=3.8$ is not compatible with any of our fit results in the external parts of the cluster.
 
The power law fits in the regions $0.7<r/r_{200}<1.2$ suffer from the fact that we a) have relatively large error bars per bin and b) do not have a large number of bins. This explains our low  $\chi^2_{red.}$ values and the large error bars for the fit parameters. Since sub sample 1 has such a high value for $\alpha$ we also show the fit result for sub sample 2, which contains all clusters of sub sample 1 with the exception of A578, the coolest one. We see a dramatic difference of the fitted $\alpha$ in sub sample 1 and 2. There is a good agreement between the fit results of sub sample 2 and sub sample 4 within the error bars. However, sub sample 1 and 4 give, as sub sample 1 and 2 results not compatible within the error bars. We also show the power law profiles fitted in the outer parts of the clusters in Fig.\ref{fig:beta} and Fig.\ref{fig:out}.  All power law profiles show a steeper slope than the beta-model. 

The error bars on the $ScEM$ profiles are very large in the external parts and it is therefore impossible to give an exact value for $\alpha$ and to estimate with certainty whether the different profiles have the same shape, e.g. whether clusters are 100\% self-similar in the external parts or not. We see very steep profiles in the outskirts. Because of the steepness of the profiles, slightly wrong temperature measurements, which shift the profiles along the x-axis can have dramatic effects on the power law fit results. Currently, we are only able to give a lower limit on $\alpha$ at radii $r>0.7 r_{200}$ which is $\alpha \le 4.3$. This lower limit is based on the fact that it is the lowest value, which is compatible with the fit results (based on the fit of sub sample 2).

\begin{table}
\caption{Results of the power law fitting. Error bars are $1\sigma$-level. The first column shows the sub sample number. The second column shows the interval in units of $r_{200}$ in which the power law is fitted.}
\label{tab:power}
\begin{tabular}{ccccccc}
\hline\hline
sa. & $r_{min}- $ & $N_{200} \times$ &  $\alpha$ & d.o.f. & $\chi^2_{red.}$ \\
& $r_{max}$ & $ 10^{-9}$ & & & \\
\hline
1 & 0.1-1.2  & $5.82^{6.35}_{5.36}$ & $2.79^{2.84}_{2.73}$ & 34 & 2.46 \\
1 & 0.1-1.0 & $6.06^{6.61}_{5.58}$ & $2.76^{2.81}_{2.70} $ & 27 & 2.71 \\
1 & 0.1-0.7 & $7.87^{8.65}_{7.15} $ & $2.56^{2.64}_{2.49}$ & 17 & 0.65\\
1 & 0.3-1.2 & $4.15^{4.77}_{3.61}$ & $3.35^{3.52}_{3.18} $ & 27 & 1.87 \\
1 & 0.7-1.2 & $0.50^{0.75}_{0.27}$ & $11.4^{14.5}_{8.71}$ & 14 & 0.47 \\
\hline
2 & 0.7-1.2 & $1.09^{1.64}_{0.56} $ & $6.99^{10.13}_{4.26} $ & 14 & 0.51 \\  
\hline
4 & 0.1-1.2 & $5.45^{5.89}_{5.02}$ & $3.03^{3.10}_{2.98}$ & 34 & 1.58 \\
4 & 0.1-1.0 & $5.75^{6.28}_{5.30}$ & $3.00^{3.06}_{2.94}$ & 27 & 1.46 \\
4 & 0.1-0.7 & $6.79^{7.55}_{6.12}$ & $2.89^{2.98}_{2.81}$ & 17 & 1.03 \\
4 & 0.3-1.2 & $3.57^{4.45}_{3.07}$ & $3.79^{4.18}_{3.42}$ & 27 & 0.80 \\
4 & 0.7-1.2 & $2.88^{3.77}_{2.05}$ & $5.73^{7.16}_{4.47}$ & 14 & 0.71 \\ 
\hline
all & 0.7-1.2 & $1.79^{2.56}_{1.13}$ & $7.22^{9.01}_{5.62}$ & 14 & 0.438 \\
\hline
\end{tabular}
\end{table}

\section{Discussion}
\label{sec:disc}

\subsection{The beta-model parameter $\beta$ as function of temperature}

We fit systematically higher $\beta$-values to the overall cluster profiles than previous studies, which find generally $\beta=2/3$. Additionally, previous work has shown an increase of $\beta$ with temperature (e.g. Castillo-Morales \& Schindler \cite{cs03}; Sanderson et al. \cite{spf}).\footnote{A dependence of $\beta$ on the temperature changes the slope of the $M-T$ relation, which we will discuss in Sec.\ref{sec:mt}.} 

Previous studies extracted and fitted individual X-ray profiles. The disadvantage of looking at profiles without averaging them before fitting is the limited statistics at the cluster outskirts. However, this part is essential to measure the right slope of the profiles. In Tab.\ref{tab:beta} we show our results of beta-model fitting in different regions of the clusters. We can see that the fitted $\beta$'s are systematically smaller in the central parts in comparison to fits going out to $r \ge r_{200}$. This is especially true for sub sample 1, which comprises the coolest clusters in our sample. (This trend is in agreement with the results of Bartelmann \ Steinmetz \cite{bs96}, who fit beta-models to the ICM of simulated clusters). We therefore suggest that the apparent  $\beta-T$ correlation is linked to the fact that the beta-model is not a good overall description of the clusters, especially for cold and small systems and that fitting only central parts of clusters gives systematically lower values for $\beta$ than overall beta-model fits. Since cooler clusters offer generally less statistics than hotter systems, due to lower luminosity and emission measure ($EM \propto \sqrt{T}$), they are generally detected out to smaller radii with respect to $r_{200}$ than their hotter counterparts. It is therefore not surprising that studies based on fitting individual clusters find a trend of increasing $\beta$ with cluster temperature. We can resume for clusters at the same redshift and with the same exposure time: the hotter the cluster, the larger the radius of detected cluster emission with respect to $r_{200}$, and the higher the corresponding fitted $\beta$-value.

\subsection{The L$_x$-T Relation}

 The central part of clusters is the region where most energy is radiated. Therefore, the central parts of clusters are responsible for the majority of the X-ray luminosity $L_X$ of clusters. Using a beta-model with $r_c=0.15 r_{200}$ and $\beta=0.8$, based on our best fit results, we calculate that 90\% of $L_X$ comes from the region $r\le 0.4 r_{200}$. We see deviations from the predicted self-similarity in regions smaller than $0.4 r_{200}$. At larger radii the $ScEM$ profiles of sub sample 1 and 4 agree well within the error bars, which suggests self-similarity in the outer cluster parts.

Previous work (Edge \& Stewart \cite{es91}; Arnaud \& Evrard \cite{ae99}; Markevitch \cite{m98}; Allen \& Fabian \cite{af98}) has shown that the observed $L_X$-T relation deviates from predictions based on simple self-similarity. Instead of 

\begin{equation}
L_X \propto T^2 (\propto M_{gas}^2 \Lambda(T)/R^3) 
\label{equ:lxtheo}
\end{equation}
, 
\begin{equation}
L_X \propto T^3
\end{equation}

with a certain error was found. It is logically to assume that the deviations from self-similarity in the central regions of clusters and the observed $L_X-T$ relation are linked together. If we assume that clusters are self-similar for $r\ge 0.4 r_{200}$ (which implies $L_X \propto T^2$ in these parts), we can calculate the needed relationship $L_X \propto T^x$ for $r \le 0.4 r_{200}$ to obtain the observed overall $L_X \propto T^3$ relation:

\begin{equation}
0.9T^x + 0.1 T^2 = T^3
\end{equation} 

this gives $x\sim 3.1-3.5$ for temperatures which vary, as in our sample between 2 and 8 keV. If our hypothesis is correct, we should observe a $L_X-T$ relation within a projected radius $r \le 0.4 r_{200}$ which scales as $L_X \propto T^{3.1...3.5}$. Assuming that $R\propto T^{1/2}$, and that the differences originate entirely from the $M_{gas}-T$ relation at $r \le 0.4 r_{200}$ with 

\begin{equation}
 M_{gas}(r \le 0.4 r_{200}) \propto T^y
\end{equation}

we find using eq.\ref{equ:lxtheo} $y = 2-2.2$ instead of  $y=3/2$, which is predicted from self-similarity. 

Our sample ranges from roughly 2 keV up to 8 keV clusters. $y=2-2.2$ implies that a 8 keV cluster has 2-2.6$\times$ more gas mass than predicted from self-similarity with respect to a 2 keV cluster($(8/2)^{2-3/2}=2, (8/2)^{2.2-3/2}=2.6 $ ). However, this is only valid for $r \le 0.4 r_{200}$. The fraction of gas mass in this region is about 6-7\% of the total gas mass calculated up to $r_{200}$. The increase of $M_{gas}$ is therefore very modest for the entire cluster.

\begin{figure}
\resizebox{\hsize}{!}{\includegraphics{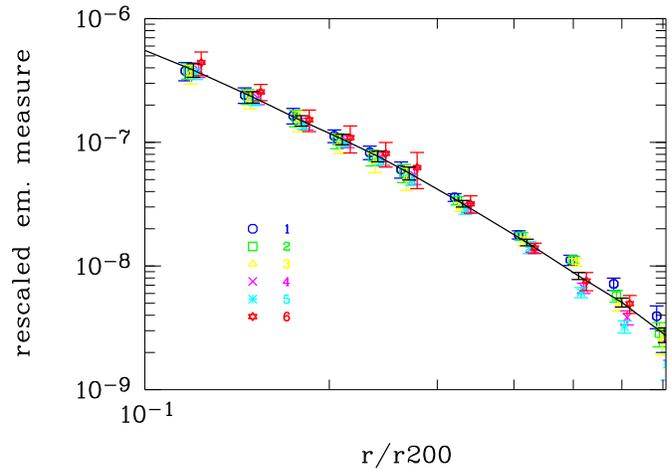}}
\caption{The different $ScEM$ profiles divided by the mean cluster temperature for each sub sample. This corresponds to $M_{gas} \propto T^2$ instead of $M \propto T^{3/2}$ predicted from simple scaling laws.}
\label{fig:logna}
\end{figure}

We can check whether this predicted $M_{gas}-T$ relation in the central region matches our data. $ScEM$ should be independent on the temperature in the self-similar case and $ScEM \propto T^{1....1.4}$ to explain the observed $L_X-T$ relation. Fig.\ref{fig:disptemp} shows $ScEM \propto T$ and $ScEM \propto T^{1.4}$ for different radii normalized to 6~keV. We see indeed, that the curves match quite well the data points in the cluster center, which reinforces the idea, that it is the observed deviation from self-similarity in the central parts of clusters, which explains the $L_X-T$ relation. In another representation we show in Fig.\ref{fig:logna} the $ScEM$ divided by the mean temperature for each sub sample. The different profiles should therefore coincide if they scale with T and the resulting gas mass with $M_{gas} \propto T^2$ instead of $M \propto T^{3/2}$, which is predicted from simple scaling laws.  As one can see at $r<0.4r_{200}$ the profiles coincide rather nicely, however, at larger radii the profiles diverge.

The idea that the deviation of the gas mass from self-similarity expectations is responsible for the observed $L_X-T$ relation is not new, we presented this idea already in another paper (Neumann \& Arnaud \cite{na01}). However, we see for the first time, that this break of self-similarity is only present in the central parts.

\subsection{Entropy in the cluster centers}

It is now common to write the entropy of the ICM in terms of  $S\propto T/n_e^{3/2}$. If clusters were perfectly self-similar the ICM density should be independent on temperature and thus $S \propto T$ should be observed. However, $S \propto T^{0.65}$ is found (Ponman et al.\cite{psf}; Pratt \& Arnaud \cite{pa03}; Pratt \& Arnaud \cite{pa05}). We observe in the central parts of the clusters in our sample that $ScEM \propto T$ instead of a $ScEM$ independent of temperature. The deviation of $S$ from self-similarity and the observed ScEM-T relation represent the same fact: $n \propto \sqrt{T}$. This result is in agreement with our previous work (Neumann \& Arnaud \cite{na01}), where we studied a sample of clusters in the z interval $0.04<z<0.06$. The deviation of the entropy from the self-similar prediction is commonly an argument for the existence of non-gravitational heating in clusters even though recently radiative cooling to explain the feature became an issue (e.g. Muanwong et al. \cite{mtkp02}).

We can synthesize the results: the observed $L_X-T$ relation of $L_X \propto T^3$, $S \propto T^{0.65}$ and $ScEM \propto T$ are caused by the fact that $n\propto \sqrt{T}$ instead of a density which is independent on the temperature in the central parts of clusters at $r<0.4 r_{200}$. 

Despite this good agreement of the work by different authors, there exist some important discrepancies: Ponman et al. (\cite{psf}) observe the entropy scaling of $S\propto T^{0.65}$ up to very large radii, which is in clear contradiction with our results, which suggest self-similarity in the outer parts ($r>0.4 r_{200}$) and thus  $S \propto T$. Ponman et al. (\cite{psf})  fit beta-models to the surface brightness profiles to individual clusters in their sample and then add together profiles of clusters in different temperature intervals\footnote{See also Sanderson \& Ponman \cite{sp} }. This is, at first sight, similar to our procedure here, however the main difference is that we add the profiles together {\it before} we fit the beta-model and {\it not after} fitting. Adding the profiles together before fitting has the advantage of enhancing the statistics of the profiles themselves, especially in our case in the outer cluster parts. These regions generally provide very low statistical quality with respect to the center. Only in adding up different profiles we have sufficient statistics to trace the X-ray profiles out to $r_{200}$ and beyond.

The differences of the $S-T$ relation in the center and in the outer cluster parts is an important piece of information for the physical modeling of the ICM. A lot of effort has been already put into explaining the observed deviation of the $S-T$ relation from self-similarity. Cavaliere et al. \cite{cmt99} developed a model in which hierarchical clustering and thus in-fall from substructures was taken into account. Tozzi \& Norman (\cite{tn01}) and Dos Santos \& Dor\'e (\cite{dd02}) looked at the effects of an initial background entropy level and shock heating. More recent models based on simulations integrate galaxy feedback and radiative cooling (Kay et al.\cite{ktjp04}; Kay \cite{k04}; Muanwong et al.\cite{mtkp02}). Voit \& Ponman (\cite{vp03}) propose smooth accretion models to explain the observed entropy scaling. For a comprehensive summary on entropy profiles and their impact on clusters as well as general cluster properties see Voit (\cite{v04}). Very recently another approach was presented by taking into account additionally AGN and quasar activity (Lapi et al. \cite{lcm05}), which yields good agreement with observations. 

The fact that we only find departures from self-similarity in the inner cluster parts suggests that the source(s) responsible for the relation $S \propto T^{0.65}$ act(s) only or mainly in the center of clusters. One can see that the $ScEM$ profiles are parallel in the centers (see Fig.\ref{fig:disp14} and Fig.\ref{fig:logna}) (see also Neumann \& Arnaud \cite{na99}; Pratt \ Arnaud \cite{pa05}). This implies that $S \propto T^{0.65}$ and $n \propto \sqrt{kT}$ not only at $r=0.1 r_{200}$ as suggested by Ponman et al. \cite{psf} but all the way up to $r=0.4 r_{200}$. The offset of the parallel curves with respect to each other defines the deviation of the $S-T$ relation from the expected simple scaling laws. The fact that the $ScEM$ profiles are parallel suggests that at each radius the deviation of the $S-T$ relation is strongly correlated to the gas density $n$. It is therefore logical to assume that it is a process which depends on $n$ which is responsible for the observed $S-T$ relation. A natural process in this respect would be radiative cooling. Radiative cooling could also explain why we only see the departures from self-similarity in the center. In the outer parts the density is not high enough to make this mechanism sufficiently efficient to be observed.

\subsection{The $M_{200}-T$ relation}
\label{sec:mt}

Making use of the beta-model and taking eq.\ref{equ:mlarge} and eq.\ref{equ:m200} together, we can calculate the $r_{200}-T$ and $M_{200}-T$ relation at any given redshift:

\begin{equation}
r_{200}(z) = \frac{0.89}{E(z)} {\sqrt{{kT}\over{\mbox{keV}}}} \mbox{Mpc}
\end{equation}

and 

\begin{equation}
M_{200}(z) = \frac{8.0 \times 10^{13}}{E(z)} \left({{kT}\over{\mbox{keV}}}\right)^{3/2} M_\odot
\end{equation}

Please keep in mind that these are the relations assuming that the ICM is isothermal. 

Bryan \& Norman (\cite{bn98}) calculated a $M-T$ relation of clusters based on Eulerian hydrodynamic simulations. Using our choice of cosmological parameters they found for an over-density $\Delta_c E^2(z)=101$ at z=0:

\begin{equation}
M = 12.4 \times 10^{13} \left(\frac{kT}{\mbox{keV}}\right)^{3/2} M_\odot	\end{equation}

We can calculate our $M-T$ relation for the same over-density (multiplying the $M_{200}-T$ relation based on the beta-model with $\sqrt{200/101}$), which gives:

\begin{equation}
M = 11.2 \times 10^{13} \left(\frac{kT}{\mbox{keV}}\right)^{3/2} M_\odot	
\end{equation}

We see a difference of 10\% with respect to the results of Bryan \& Norman (\cite{bn98}). The corresponding observed $R-T$ relation has a 3\% lower normalization than the relation based on simulations. This difference is quite small and is less than found in other studies based on X-ray data, which give normalisations much lower than the results of Bryan \& Norman (\cite{bn98}) (e.g. Horner et al. \cite{hms}; Mohr et al. \cite{mme99}; Finoguenov et al. \cite{frb}; Nevalainen et al. \cite{nmf}; Allen et al.\cite{asf01}; Arnaud et al.\cite{app05}). \footnote{Please note: when the $M-T$ relation is not given for $M_{200}$, we extrapolated this relation assuming a constant density gradient and isothermality}
 The difference between our results and previous studies is the fact that we see a steeper slope for the profile in outer parts, which translates in a higher value for $\beta$. Using the hydrostatic equation, the assumption of isothermality of the ICM  and the beta-model to calculate the $M-T$ relation one finds that the normalization of this relation is directly proportional to $\beta$. Our results of a steepening of the $ScEM$ profiles is in agreement with a study of Vikhlinin et al. (\cite{vfj}). These authors fitted beta-models in the outer parts of clusters and found systematically higher $\beta$'s with respect to fits performed taking the cluster centers into account.

We have seen that the slope of the $ScEM$ profile in the outer parts is quite likely steeper than the beta-model with $\beta=0.80$. We can calculate the ICM density profile and subsequently the $M_{200}-T$ relation taking the power law model (eq.\ref{equ:power}) used to fit the surface brightness profile: 

\begin{equation}
n(r) \propto r^{-(\alpha+1)/2}
\label{equ:icmrho}
\end{equation}

\begin{equation}
M_{200} = 7.4 \times 10^{12} \times (\alpha+1)^{3/2} \left(\frac{kT}{\mbox{keV}}\right)^{3/2}M_\odot
\label{equ:mpower}
\end{equation} 

As one can easily see, a higher value for $\alpha$ translates in a higher normalization of the $M-T$ relation.

We observe that the $ScEM$ profile drops more rapidly than the beta-model with $\beta=0.8$ at radii larger than $0.8 r_{200}$. Power law fits in the outer parts give a power law index $\alpha \ge 4.3$. Taking $\alpha=4.3$ we find a normalization of the $M-T$ relation of 12.6 (see eq.\ref{equ:mpower}), which is in very good agreement with the value of 12.4 found by Bryan \& Norman (\cite{bn98}). We will come back on the issue of the slope in the next section.

\subsection{Indications for instabilities at large radii}

Recently,  analyzing a sample of nearby clusters observed with XMM-Newton Pointecouteau et al. \cite{pap} (see also Pointecouteau et al.\cite{pakd04}) have found, that the matter profiles in clusters fit very well the NFW-profile (Navarro, Frank \& White \cite{nfw96}; \cite{nfw97}), which has the density distribution 

\begin{equation}
\rho(r) = \frac{\rho_0}{(r/r_s)(1+r/r_s)^2}
\end{equation}
 
The results found by Pointecouteau match results based on Chandra data (Arabadjis et al. \cite{abg02}, \cite{aba04}; Buote \& Lewis \cite{bl}; Lewis et al.\cite{lbs}).

\begin{figure}
\resizebox{\hsize}{!}{\includegraphics{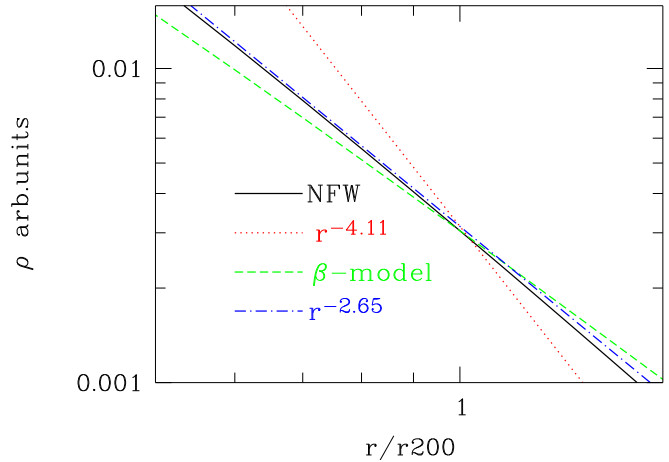}}
\caption{Different density profiles in arbitrary units normalized to the same value at $r=r_{200}$. Since the NFW profile and the profile representing $r^{-2.65}$ have such an identical shape we choose a small offset between the two curves at $r_{200}$ to enhance the visibility of the plot. The line $r^{-2.65}$ represents the shallowest density profile, which is in agreement with our power law fits on the $ScEM$ profiles in the external cluster parts. The curve with  $r^{-4.11}$ represents the density distribution which originates from the best power law fit to the $ScEM$ profile averaged over all clusters in the outer regions with $ScEM(r) \propto r^{-7.22}$ (see also Tab.\ref{tab:power}).}
\label{fig:nfw}
\end{figure}

There is general agreement between observations and simulations (Dolag et al. \cite{dolag}) that the concentration parameter $c = r_{200}/r_s$ is approximately 6, which implies that $r_s=0.16-0.17 r_{200}$. We show in Fig.\ref{fig:nfw} the density distribution of the NFW-profile, assuming $r_s=0.16$ and compare it with a beta-model slope with $\beta=0.8$, and a slope with $r^{-4.11}$, which corresponds to the density profile inferred from the best fit power law in the outskirts averaged over all $ScEM$ profiles (see Tab.\ref{tab:power}). We also show in Fig.\ref{fig:nfw} $r^{-2.65}$, which corresponds to the density fall-off of the ICM if $S=r^{-\alpha}$, with $\alpha=4.3$, which is our lower limit at $r_{200}$  (see also Sec.\ref{sec:mt}). As one can see in this figure, the slope of $r^{-2.65}$ is parallel to the NFW density profile around $r_{200}$. Therefore it is not unlikely that the ICM density falls off more rapidly close to $r_{200}$ than predicted from the NFW profile for cold dark matter. How can we interpret this result? 

At some radius in the outskirts of clusters a strong temperature gradient must occur, since we know from COBE results (Smoot et a.\cite{cobe}) that the material surrounding a cluster has a temperature of roughly 3~K. The ICM drops from a few $10^7-10^8$~K down to a few K within a short radial range. This large temperature gradient makes the ICM convectively unstable and as a result the cold material from the outskirts  and the hot ICM mix. This mixing is enhanced by accretion of cold material from the cluster outskirts. This very likely causes that the ICM at large radii is not anymore in hydrostatic equilibrium. It is very likely that our result reflects this scenario. We only see the hot ICM emitting in X-rays, but we do not see cooler material or in general gas, which is below a temperature of about $10^7$~K. It is therefore likely that the total baryon fraction in the outskirts is similar to the one observed closer to the cluster center. However, at larger and larger radii the hot phase becomes less and less important and cooler components not emitting in X-rays contribute more and more to the total baryon budget. It is currently not clear how the different phases mix in detail and how effective heat conduction is in this respect. More detailed studies on cluster outskirts similar to the study presented here are necessary to give quantitative numbers. For this it is important to have the temperature information of the ICM at radii $>>r_{500}$. Unfortunately, this is currently very difficult to obtain. One would need a very high throughput X-ray observatory with good energy resolution, low instrumental background and large field-of-view (FOV) to do this kind of study. XMM-Newton and Chandra unfortunately do not have sufficiently low internal background at high energies and provide only a limited FOV to fulfill these requirements. Maybe more distant clusters which fit in the FOV of XMM-Newton and Chandra can be used for a stacking method to constrain the temperature distribution of the ICM at large radii.

\section{Summary and conclusion}

In the study presented here we looked at a sample of 14 nearby clusters of galaxies in the redshift range $0.06<z<0.1$ with low galactic absorption for which good temperature measurements are available in the literature. For each cluster we calculate the emission measure profile scaled to $r_{200}$ and apply the spherical collapse model  (Eke et al. \cite {eke96} and Bryan \& Norman \cite{bn98}). We divide the clusters in sub samples which represent different temperature intervals. In order to enhance the statistics, especially in the cluster outskirts, we stack the profiles of clusters belonging to the same sub sample together. Comparing the different scaled and stacked emission measure (ScEM) profiles we find the following results:

\begin{enumerate}

\item The scaled emission measure profiles are not self-similar in the center, however, in a logarithmic representation the different profiles are parallel in the range $0.1 r_{200}<r<0.4 r_{200}$. The difference of the profiles can be approximated with $ScEM \propto  T$ which implies a gas density $n \propto \sqrt{T}$. This is in agreement with entropy studies, which find generally $S\propto T^{0.65}$ (Ponman et al.\cite{psf}; Pratt \& Arnaud \cite{pa03}, \cite{pa05}). The fact that the different $ScEM$ profiles are parallel up to $r<0.4 r_{200}$ suggests that the difference of the $S-T$ relation, which defines the offset of the profiles with respect to each other is strongly coupled to the ICM density itself. The only currently recognized mechanism  which links the ICM density to the observed $S-T$ relation is radiative cooling.

\item We find that about 90\% of the X-ray emission of clusters comes from the region encompassing $r<0.4 r_{200}$. Therefore this central region is primarily responsible for the observed X-ray luminosity. The deviation of the $ScEM$ profiles from self-similarity in the centre which is responsible for the observed $S-T$ relation explains also the observed $L_X-T$ relation $L_X\propto T^3$ (Edge \& Stewart \cite{es91}; Allen \& Fabian \cite{af98}; Markevitch \cite{m98}; Arnaud \& Evrard \cite{ae99}), which deviates from simple expectations based on structure formation and gravity.

\item At radii $r>0.4 r_{200}$ the different cluster profiles agree well within the error bars, which suggests self-similarity. This implies that global cluster properties defined at large radii, such as the $M-T$ or $R-T$ relation follow the theoretical scaling of $M\propto T^{3/2}$ and $R\propto \sqrt{T}$. 

\item Fitting beta-models to the overall stacked profiles we find generally $\beta=0.8$, which is higher than the canonical value of $\beta=2/3$, which is often observed. The found higher value of $\beta$ translates into the need of a  higher normalization of the $M-T$ relation with respect to previous studies. Assuming isothermality up to large radii we find a normalization of the $M-T$ which is only 10\% lower than the relation based on numerical simulations found by Bryan \& Norman (\cite{bn98}). 

The beta-model is generally a better representation of the hotter than of the cooler clusters. We find relatively poor agreement between the beta-model and the $ScEM$ profiles in the central parts ($r \le 0.15 r_{200}$) of cool clusters. Simple scaling laws predict that the emission measure scales with $EM \propto \sqrt{T}$. This effect causes that cooler clusters have systematically smaller detection radii with respect to $r_{200}$ than hotter clusters. The bad representation of the X-ray profiles of the central parts of cool clusters and $EM \propto \sqrt{T}$ are quite likely the reasons why previous studies looking at individual cluster profiles found a correlation of increasing $\beta$ with increasing temperature. 

\item The $ScEM$ profiles are systematically below the best-fit beta-model at radii $r>0.8 r_{200}$, which implies a steepening of the X-ray profiles with radius. Fitting power laws to the external parts of the $ScEM$ profiles, we find steeper profiles than predicted from the NFW-profile, which fits observed cluster density profiles well up to $r_{500}$. We argue that this steepening of the observed profiles is caused by the fact that at radii close to $r_{200}$ the ICM is not anymore in hydrostatic equilibrium, as predicted from numerical simulations. At these large radii the X-ray emitting ICM is mixed very likely with  cooler material accreted from larger radii. This multiphase gas causes that part of the baryons are not anymore detected in X-rays. 

\end{enumerate}

Our cluster sample spans over the temperature range $1.7$keV$<kT<8.5$keV, however only one cluster has a temperature below 2~keV and only one cluster is above 8~keV. Therefore our dynamic range in temperature is limited. We can therefore currently not exclude that the behavior of small groups with temperatures well below 1~keV or above 9~keV is in agreement with our results or not.

\begin{acknowledgements}
The author would like to thank to the anonymous referee for constructive comments and fast replies.

The author wishes is greatful to M. Arnaud, R. Teyssier, F. Motte  and E. Pointecouteau for useful discussions and is thankful to Y. and I. Maazi, W. Neumann and M.S. Schweinderle for their patience and support. The author would also like to thank M.-P. Genebrias for stimulating advice.

This research has made use of the NASA/IPAC Extragalactic Database (NED) which is operated by the Jet Propulsion Laboratory, California Institute of Technology, under contract with the National Aeronautics and Space Administration. This research has made also use of the X-Rays Clusters Database (BAX) which is operated by the Laboratoire d'Astrophysique de Tarbes-Toulouse (LATT), under contract with the Centre National d'Etudes Spatiales (CNES). Also: this research has made use of NASA's Astrophysics Data System.

\end{acknowledgements}

\end{document}